\documentstyle[preprint,aps]{revtex}
\begin{document}
\preprint{HYUPT 0825/96}
\draft
\title{Conformal Couplings in Induced Gravity \footnote{To be appeared 
in General Relativity and Gravitation}}
\author{C.J. Park$^{+}$ and Yongsung Yoon$^{\dagger}$}
\address{$^{+}$Department of Physics, Seoul National University \\
Seoul 151-742, Korea \\
$^{\dagger}$Department of Physics, Hanyang University 
Seoul 133-791, Korea
}
\maketitle
\begin{abstract}
It is found that the induced gravity with conformal couplings requires 
the conformal invariance in both classical and quantum levels for consistency.
This is also true for the induced gravity with an extended conformal coupling
interacting with torsion.
\end{abstract}

\section{Introduction}

Far below the electro-weak scale, the weak interaction is well 
characterized by the dimensional Fermi's coupling constant,
$G_{F}= (300 Gev)^{-2}$.
However, from the success of Weinberg-Salam model,
it turns out that the dimensional coupling constant is the low
energy effective coupling which is determined by the dimensionless electro-weak 
coupling constants and the vacuum expectation value of Higgs scalar field
through the spontaneous symmetry breaking.
Indeed $G_{F}\cong v_{\omega}^{-2}$, where $ v_{\omega}\cong 300$ Gev
is the vacuum expectation value of Higgs field.
The weakness of the weak interaction comes from the largeness of 
the vacuum expectation value of Higgs field \cite{ref1}. 
Thus, among the four fundamental interactions in nature, only gravitational
interaction is characterized by the dimensional coupling constant,
Newton's constant $G_{N}\cong (10^{19} Gev)^{-2}$.

It's well known that the interactions with dimensional coupling constants of
inverse mass dimension are strongly diverse and nonrenomalizable.  From the 
success of the Weinberg-Salam model, it might be considerable that 
gravity is also characterized by a dimensionless coupling constant $\xi$, and
that the weakness of gravity is associated with a symmetry breaking
at high energy. Similarly to $G_{F}$, 
$G_{N}$ could be given by the inverse square of the vacuum expectation  
value of a scalar field, dilaton. 
It was independently proposed by Zee \cite{ref2}, Smolin \cite{ref3},
and Adler \cite{ref4} that the Einstein-Hilbert action can be replaced by
the induced gravity action
\begin{equation}
     {\cal S} =  \int d^{4}x \sqrt{g}(\frac{1}{2}\xi \phi^{2} R +
	  \frac{1}{2} \partial_{\mu}\phi \partial^{\mu}\phi - V(\phi)),
\label{action}
\end{equation}
where the coupling constant $\xi$ is dimensionless.  
The potential $V(\phi )$ is
assumed to attain its minimum value when  $\phi = \sigma$, then
$G_{N}=\frac{1}{8\pi\xi\sigma^{2}}$.

On the analogy of the $SU(2) \times U(1)$ symmetry of the 
electro-weak interactions, we can consider a symmetry which is broken through
spontaneous symmetry breaking in the gravitational interactions.  
Through the spontaneous symmetry breaking, the symmetric  phase of the 
scalar field transits to an asymmetric phase of the scalar field.
There have been several attempts to apply some spontaneous symmetry
breakings in induced gravity to inflationary models
\cite{ref5,ref5a,ref6,ref7,ref8}.
One of the most attractive symmetry in induced gravity is the conformal
symmetry which rejects the Einstein-Hilbert action, but admits the
induced gravity action Eq.(\ref{action}) with the specific conformal coupling
$\xi = \frac{1}{6}$.

In Riemann-Cartan space, the vector torsion plays the role of the 
conformal gauge field \cite{ref9,ref9a}.
Without the vector torsion, the conformally invariant induced gravity
action is unique with the specific conformal coupling.
However, introducing the vector torsion field, a conformally invariant
extension of induced gravity action can be considered \cite{ref10}.

Actually, we don't know yet whether nature really shows conformal invariance
at a sufficiently high energy scale.
But there is some evidences for this conformal
invariance from the renormalization group analysis of some induced
gravity models. For some $SU(N)$ induced gravity models, it is found that,
at high energy limit, the coupling $\xi$ approaches to the conformal 
coupling \cite{ref11,ref12,ref13}.
If all other interactions including scalar potential are conformally
invariant in this limit, then the models show asymptotic conformal invariance.
This may happen also for some Grand Unified Models with induced gravity action
\cite{ref14}.

We have investigated the conformal couplings in induced gravity 
and found that induced gravity at conformal couplings should have conformal
invariance for consistency. An extension of conformal coupling in induced 
gravity is also considered introducing the vector torsion.

\section{Extension of Conformal Coupling in Induced Gravity} 

In this section, we consider an extension of conformal coupling 
introducing the torsion in induced gravity action.
The induced gravity action Eq.(\ref{action}) is invariant under the 
conformal transformation,
\begin{equation}
    g_{\mu\nu}'(x) = exp(2\Lambda) g_{\mu\nu}(x), ~~~
    \phi'(x) = exp(-\Lambda) \phi(x),
\label{tran1}
\end{equation}
at the conformal coupling $\xi = \frac{1}{6}$ for a conformally invariant
scalar potential.

To consider an extension of conformal coupling with the torsion in induced
gravity, we have to introduce Riemann-Cartan space-time first.
However, it is found that the minimal extension to Riemann-Cartan space-time
is sufficient.
The conformal transformation of the affine connections
$\Gamma^{\gamma}_{~\beta \alpha}$ is determined from the 
invariance of the tetrad postulation,
\begin{equation}
       D_{\alpha} e^{i}_{\beta}\equiv\partial_{\alpha}e^{i}_{\beta} +
       \omega^{i}_{j\alpha} e^{j}_{\beta}-
       \Gamma^{\gamma}_{~\beta \alpha}e^{i}_{\gamma} = 0,
\label{cor}
\end{equation}
under the following tetrads $e^{i}_{\alpha}$ and the spin connections
$\omega^{i}_{j\alpha}$ transformations;
\begin{equation}
	(e^{i}_{\alpha})'= exp(\Lambda) e^{i}_{\alpha}, ~~~
	(\omega^{i}_{j\alpha})' = \omega^{i}_{j\alpha}.
\label{tran2}
\end{equation}
We have used Latin indices for the tangent space-time and  Greek indices  
for the curved space-time. 
From the metric compatibility Eq.(3), the affine connections and the torsions 
which are the antisymmetric components of the affine connections transform 
as follows;
\begin{equation}
    (\Gamma^{\gamma}_{~\beta \alpha})'
    =\Gamma^{\gamma}_{~\beta \alpha} + \delta^{\gamma}_{\beta} \partial_{\alpha}
    \Lambda, ~~~
    (T^{\gamma}_{~\beta \alpha})'=
    T^{\gamma}_{~\beta \alpha}+
    \delta^{\gamma}_{\beta}\partial_{\alpha}\Lambda
    - \delta^{\gamma}_{\alpha}\partial_{\beta}\Lambda.
\label{tor1}
\end{equation}
Therefore, the contracted vector torsion $T^{\gamma}_{~\gamma \alpha}$ is 
effectively playing the role of a conformal gauge field. 
The torsion tensor can be decomposed into three irreducible components
\cite{ref17}.
However, it is sufficient for the purpose of examination to decompose the
torsion into the conformally invariant tracefree tensor part 
$A^{\alpha}_{~\beta \gamma}$ and conformally non-invariant vector part
$S_{\alpha}$ as follows;
\begin{equation}
          T^{\alpha}_{~\beta \gamma}=
          A^{\alpha}_{~\beta \gamma}-
	  \delta^{\alpha}_{\gamma}S_{\beta}+
	  \delta^{\alpha}_{\beta}S_{\gamma},
\label{tor2}
\end{equation}
\begin{equation}
          (S_{\alpha})' =
	  S_{\alpha}+
	  \partial_{\alpha}\Lambda, ~~~
          (A^{\alpha}_{~\beta \gamma})'=
          A^{\alpha}_{~\beta \gamma} .
\label{vec}
\end{equation}
This conformal transformations Eq.(4) and Eq.(7) are also considered in 
\cite{ref3,ref18} to construct conformally invariant Ricci tensor.

Because minimal extension to Riemann-Cartan space-time is sufficient,
we impose the conformally invariant torsionless condition 
\begin{equation}
          A^{\alpha}_{~\beta \gamma} \equiv 0.
\label{zero}
\end{equation}
This  condition is the conformally invariant extension of the
torsionless condition in Riemann space-time
$ T^{\alpha}_{~\beta \gamma}\equiv 0$.
For this space, the affine connection can be written in terms of 
$g_{\mu\nu}$ and $S_{\alpha}$;
\begin{equation}
          \Gamma^{\alpha}_{~\beta \gamma}=
	  \{^{\alpha}_{\beta \gamma}\}+
          S^{\alpha} g_{\beta\gamma}-
          S_{\beta}\delta^{\alpha}_{\gamma}.
\label{cristo}
\end{equation}
Defining the conformally invariant connection 
$\Omega^{\alpha}_{~\beta \gamma}$,
\begin{equation}
          \Omega^{\alpha}_{~\beta \gamma} \equiv
	  \{^{\alpha}_{\beta \gamma}\}+
          S^{\alpha}g_{\beta\gamma}-
          S_{\beta}\delta^{\alpha}_{ \gamma}-
          S_{\gamma}\delta^{\alpha}_{ \beta},
\label{con1}
\end{equation}
the curvature tensors $R^{\alpha}_{~\beta \mu \nu}(\Gamma)$
of the affine connections
can be expressed in terms of
the curvature tensors $R^{\alpha}_{~\beta \mu \nu}(\Omega)$ of
$\Omega^{\alpha}_{~\beta\gamma}$ and $S_{\alpha}$;
\begin{equation}
R^{\alpha}_{~\beta \mu \nu}(\Gamma)=
R^{\alpha}_{~\beta \mu \nu}(\Omega)+
\delta^{\alpha}_{\beta}H_{\mu \nu}, ~~~
R_{\alpha \nu}(\Gamma)=R_{\alpha \nu}(\Omega)+H_{\alpha \nu},
\label{curv}
\end{equation}
where $H_{\mu \nu} \equiv \partial_{\mu}S_{\nu}- \partial_{\nu}S_{\mu}$
is the field strength of the vector torsion $S_{\alpha}$.
With the help of Eq.(\ref{cristo}) and Eq.(\ref{curv}),
we obtain the identity,
\begin{equation}
\sqrt{g}R(\Omega)=\sqrt{g}R(\{\}) +
6\sqrt{g}(\nabla_{\alpha}S^{\alpha}-S_{\alpha}S^{\alpha}),
\label{ident}
\end{equation}
where $\nabla_{\alpha}$ is the ordinary covariant derivative
in Riemann space-time.

Introducing the conformally covariant derivative $D_{\alpha}$,
\begin{equation}
D_{\alpha}\phi \equiv \partial_{\alpha}\phi + S_{\alpha}\phi,
\end{equation}
we have an extended conformal coupling in induced gravity
up to total derivatives as follow;
\begin{equation}
      {\cal S} =\int d^{4}x \sqrt{g}
      (\frac{\xi}{2}R(\Omega)\phi^{2} +
 	\frac{1}{2}D_{\alpha}\phi D^{\alpha}\phi -
 	\frac{1}{4}H_{\alpha\beta}H^{\alpha\beta} -
 	V(\phi)),
\label{gen}
\end{equation}
where we have excluded the curvature square terms and conformally
non-invariant torsion terms like $S^{\mu}S_{\mu}$.
Now, the coupling $\xi$ is a dimensionless arbitrary constant.
Using Eq.(\ref{ident}) we can rewrite this action  in
terms of Riemann curvature scalar $R(\{\})$;
\[
      {\cal S}=\int d^{4}x \sqrt{g}
      (\frac{\xi}{2}R(\{\})\phi^{2} +
 	\frac{1}{2}\partial_{\alpha}\phi\partial^{\alpha}\phi-
 	\frac{1}{4}H_{\alpha \beta}H^{\alpha \beta} 
\]
\begin{equation}
	~~~~~~+(1-6\xi) S^{\alpha}(\partial_{\alpha}\phi)\phi+
 	\frac{1}{2}(1-6\xi)S_{\alpha}S^{\alpha}\phi^{2}-
 	V(\phi)).
\label{iaction}
\end{equation}
In the limit of $\xi \rightarrow \frac{1}{6}$, this extended conformal
coupling is reduced to the ordinary conformal coupling in Riemann space-time
without torsion.

\section{Conformal Invariance at Conformal Couplings}

We analyze the equations of motion for the action Eq.(\ref{iaction}),
in which the potential $V(\phi)$ does not need to be classical,
but can be an effective scalar potential $V_{eff}(\phi)$ after
integrating out all fluctuating quantum fields.
Moreover, in general, the effective scalar potential may depend on 
the other background fields, the metric $g_{\mu\nu}$ and the vector
torsion $S_{\alpha}$. Thus, we will consider the potential in the
action Eq.(\ref{iaction}) as an effective scalar potential
$V_{eff}(\phi;g_{\mu\nu},S_{\alpha})$.

Varying the action, we obtain the three equations of motion;
\begin{equation}
\Box\phi=
\xi R(\{\})\phi
+(1-6\xi)\phi(S^{\mu}S_{\mu}-\nabla_{\mu}S^{\mu})
-\frac{\partial V_{eff}(\phi;S_{\alpha},g_{\beta\gamma})}{\partial\phi}~,
\label{box}
\end{equation}
\begin{equation}
\partial_{\mu}(\sqrt{g}H^{\mu\nu})=
-(1-6\xi)\sqrt{g}\{(\partial^{\nu}\phi)\phi+S^{\nu}\phi^{2}\}
+\frac{\partial V_{eff}(\phi;S_{\alpha},g_{\beta\gamma})}{S_{\nu}}~,
\label{box1}
\end{equation}
\[
\xi\phi^{2}G_{\mu\nu}=
(H_{\mu\alpha}H_{\nu}^{~\alpha}
  -\frac{1}{4}g_{\mu\nu}H_{\alpha\beta}H^{\alpha\beta})
-(\partial_{\mu}\phi\partial_{\nu}\phi 
  -\frac{1}{2} g_{\mu\nu}\partial_{\alpha}\phi\partial^{\alpha}\phi)
-(1-6\xi)\phi^{2}(S_{\mu}S_{\nu}-\frac{1}{2}g_{\mu\nu}S_{\alpha}S^{\alpha})
\]
\[
-(1-6\xi)(S_{\mu}\phi\partial_{\nu}\phi+S_{\nu}\phi\partial_{\mu}\phi
-g_{\mu\nu}S^{\alpha}\phi\partial_{\alpha}\phi)
+\xi\{\nabla_{\mu}(\phi\partial_{\nu}\phi)+\nabla_{\nu}(\phi\partial_{\mu}\phi)
-g_{\mu\nu}\Box\phi^{2}\}
\]
\begin{equation}
-g_{\mu\nu}V_{eff}(\phi;S_{\alpha},g_{\beta\gamma})
+2\frac{\partial V_{eff}(\phi;S_{\alpha},g_{\beta\gamma})}
{\partial g^{\mu\nu}}~.
\label{long}
\end{equation}
Taking the divergence of Eq.(\ref{box1}), we obtain 
\begin{equation}
(1-6\xi)\nabla_{\mu}(S^{\mu}\phi^{2})= -\frac{1}{2}(1-6\xi)\Box\phi^{2}
+\nabla_{\nu}\frac{\partial V_{eff}(\phi;S_{\alpha},g_{\beta\gamma})}
{\partial S_{\nu}}~.
\label{tran}
\end{equation}
The trace of Einstein Eq.(\ref{long}) is
\[
\xi R(\{\})\phi^{2}=
-\partial_{\alpha}\phi\partial^{\alpha}\phi
-(1-6\xi)(S^{\alpha}\partial_{\alpha}\phi^{2}+S_{\alpha}S^{\alpha}\phi^{2}) 
+3\xi\Box\phi^{2}
\]
\begin{equation}
+4V_{eff}(\phi;S_{\alpha},g_{\beta\gamma})
-2\frac{\partial V_{eff}(\phi;S_{\alpha},g_{\beta\gamma})}
{\partial g^{\mu\nu}}g^{\mu\nu}~.
\label{tra}
\end{equation}
 From Eq.(\ref{box}) and Eq.(\ref{tra}), we have 
\[
\phi\Box\phi+\partial_{\alpha}\phi\partial^{\alpha}\phi
+(1-6\xi)\nabla_{\alpha}(S^{\alpha}\phi^{2})-3\xi\Box\phi^{2}=
\]
\begin{equation}
4V_{eff}(\phi;S_{\alpha},g_{\beta\gamma})
-\phi\frac{\partial V_{eff}(\phi;S_{\alpha},g_{\beta\gamma})}{\partial\phi}
-2\frac{\partial V_{eff}(\phi;S_{\alpha},g_{\beta\gamma})}
{\partial g^{\mu\nu}}g^{\mu\nu}~.
\label{combi}
\end{equation}
Using Eq.(\ref{tran}), we have a $\xi$ independent equation for a general
effective potential from Eq.(\ref{combi}) as follows;
\begin{equation}
4V_{eff}(\phi;S_{\alpha},g_{\beta\gamma})
-\phi\frac{\partial V_{eff}(\phi;S_{\alpha},g_{\beta\gamma})}{\partial\phi}
=2\frac{\partial V_{eff}(\phi;S_{\alpha},g_{\beta\gamma})}
{\partial g^{\mu\nu}}g^{\mu\nu}
+\nabla_{\nu}\frac{\partial V_{eff}(\phi;S_{\alpha},g_{\beta\gamma})}
{\partial S_{\nu}}~.
\label{combix}
\end{equation}
Therefore the metric and vector torsion dependencies of an effective 
potential are directly related
to the deviation of the effective potential from the quartic form.

Let's consider the conformal transformations Eq.(\ref{tran1})
and Eq.(\ref{tran2}) of the action Eq.(\ref{iaction}),
in which the scalar potential is replaced by the effective potential
$V_{eff}(\phi;g_{\mu\nu},S_{\alpha})$.
Because the kinetic terms are conformally invariant for the conformal
couplings, only the scalar potential term contributes to the conformal
variation;
\[
\delta {\cal S} = \int d^{4}x \sqrt{g} ( - \frac{1}{2}
V_{eff}(\phi;g_{\mu\nu},S_{\alpha})
g_{\mu\nu}\delta g^{\mu\nu} 
+ \frac{\partial V_{eff}(\phi;g_{\mu\nu},S_{\alpha})}
{\partial \phi} \delta \phi
\]
\begin{equation}
+ \frac{\partial V_{eff}(\phi;g_{\mu\nu},S_{\alpha})}
{\partial g^{\mu\nu}}\delta g^{\mu\nu}
+ \frac{\partial V_{eff}(\phi;g_{\mu\nu},S_{\alpha})}
{\partial S_{\alpha}}\delta S_{\alpha}).
\end{equation}
Using the infinitesimal forms of the conformal transformations
Eq.(\ref{tran1}) and Eq.(\ref{tran2}),
\begin{equation}
\delta g^{\mu\nu} = -2\Lambda g^{\mu\nu}, ~~~\delta \phi = - \Lambda \phi,
~~~\delta S_{\alpha} = \partial_{\alpha} \Lambda,
\end{equation}
the conformal variation of the action can be written as
\[
\delta {\cal S} = \int d^{4}x \sqrt{g}
\Lambda (4V_{eff}(\phi;g_{\mu\nu},S_{\alpha})
- \frac{\partial V_{eff}(\phi;g_{\mu\nu},S_{\alpha})}{\partial \phi} \phi
- 2 \frac{\partial V_{eff}(\phi;g_{\mu\nu},S_{\alpha})}
{\partial g^{\mu\nu}} g^{\mu\nu}
\]
\begin{equation}
- \nabla_{\alpha} \frac{\partial V_{eff}(\phi;g_{\mu\nu},S_{\alpha})}
{\partial S_{\alpha}}) 
- \int d^{4}x \partial_{\alpha}(\sqrt{g} \Lambda 
 \frac{\partial V_{eff}(\phi;g_{\mu\nu},S_{\alpha})}{\partial S_{\alpha}}).
\end{equation}
Because the last total derivative term can be eliminated if we consider a
conformal transformation which have a vanishing $\Lambda(x)$ at space-time
infinity, the Eq.(\ref{combix}) we have obtained from the equations of 
the motion analysis is the condition for the the conformal invariance of 
the induced gravity action, $\delta {\cal S} \equiv 0$.
The relation Eq.(\ref{combix}) also appears in case of the special 
conformal coupling $\xi = \frac{1}{6}$, where the vector torsion is 
decoupled from
the scalar field. Therefore, we can say that the conformal couplings
in induced gravity generally requires the conformal invariance of the
induced gravity action for consistency.

\section{Conclusion}

Without introducing the vector torsion, the conformal coupling in induced
gravity is unique with $\xi=\frac{1}{6}$. However, in Riemann-Cartan 
space-time the vector torsions play the role of the conformal gauge fields,
which make an extended conformal coupling possible \cite{ref8}.

For some $SU(N)$ induced gravity models, it is found that the coupling $\xi$ 
approaches to the conformal coupling $\frac{1}{6}$ at high energy limit
\cite{ref11,ref12,ref13}.
If all other interactions are conformally invariant in this limit, then the
models have asymptotic conformal invariance. This may happen also for some
Grand Unified Models with induced gravity action \cite{ref14}.

We have investigated the conformal couplings in induced gravity and found that
the induced gravity models at conformal couplings should have conformal
invariance for consistency at classical and quantum levels.

~\\~\noindent
{\it Acknowledgments:} 
C.J. Park was supported by KOSEF Post-Doc Fellowship. Y. Yoon was supported
by KRF/BSRI-2441, KOSEF/94-1400-04-01-3, and Hanyang University.


\begin{references}

\bibitem{ref1}S. Weinberg, Phys. Rev. Lett. {\bf 19} ,1264 (1979).
\bibitem{ref2}A. Zee, Phys. Rev. Lett. {\bf 42}, 417 (1979).
\bibitem{ref3}L. Smolin, Nucl. Phys. {\bf B160}, 253 (1979).
\bibitem{ref4}S.L. Adler, Rev. Mod. Phys. {\bf 54}, 729 (1982).
\bibitem{ref5}F.S. Accetta, D.J. Zoller, and M. Turner Phys. Rev. D
{\bf 31}, 3046 (1985).
\bibitem{ref5a}A. Albrecht and P.J. Steinhardt, Phys. Rev. Lett. 
{\bf 48}, 1220 (1982).
\bibitem{ref6}J.L. Cervantes-Cota and H. Dehnen, Phys. Rev. D {\bf 51},
395 (1995); Nucl. Phys. {\bf B442}, 391 (1995).
\bibitem{ref7}D.I. Kaiser, Phys. Rev. D {\bf 49}, 6347 (1994);
"Induced-gravity Inflation and the Density Perturbation Spectrum",
Report No. HUTP-94/A011 (astro-ph/9405029) (unpublished).
\bibitem{ref8}J. Kim, C.J. Park, and Y. Yoon, Phys. Rev. D {\bf 51},
562 (1995).
\bibitem{ref9}H.T. Nieh and M.L. Yan, Ann. Phys. (N.Y.) {\bf 138}, 237 (1982).
\bibitem{ref9a}J.K. Kim and Y. Yoon, Phys. Lett. B {\bf 214}, 98 (1988).
\bibitem{ref10}J. Kim, C.J. Park, and Y. Yoon, Phys. Rev. D {\bf 51},
4595 (1995).
\bibitem{ref11}I.L.Buchbinder, S.D. Odintsov, and I.L. Shapiro, "{\it Effective
action in Quantum Gravity} (Institute of Physics, Bristol, 1992).
\bibitem{ref12}I.L. Buchbinder and S.D. Odintsov, Izv. VUZ
Fiz. (Sov. J. Phys.) {\bf N12}, 108 (1983);
Yad. Fiz.(Sov. J. Nucl. Phys) {\bf 40}, 1338 (1984);
Lett. Nuovo Cimento {\bf 42}, 379 (1985).
\bibitem{ref13}Youngsoo Yoon and Yongsung Yoon, "Asymptotic Conformal
Invariance of SU(2) and Standard Models in Curved Space-time",
HYUPT Preprint (1996).
\bibitem{ref14}B. Geyer and S.D. Odintsov, Phys. Rev. {\bf D53}, 7321 (1996).
\bibitem{ref17}K. Hayashi, K. Nomura and T. Shirafuji, Prog. Theor. Phys. 
{\bf 84}, 1085 (1990).
\bibitem{ref18}R.T. Hammond, J. Math. Phys. {\bf 31}, 2221 (1990).

\end{references}
\end{document}